\documentclass[aps,prb,twocolumn,superscriptaddress,longbibliography]{revtex4-1}
\usepackage{amsmath}
\usepackage{amssymb}
\usepackage{graphicx}
\usepackage{braket}
\usepackage[colorlinks = true, linkcolor = red,urlcolor  = blue, citecolor = blue,anchorcolor = blue]{hyperref}
\usepackage{srcltx}

\newcommand{\nn}{\nonumber \\}
\usepackage{soul}

\newcommand{\balig}{\begin{align}}
\newcommand{\ealig}{\end{align}}
\newcommand{\peq}{P_{N-1}^{\rm{eq}}}

\newcommand{\bee}{\begin{eqnarray}}
\newcommand{\ee}{\end{eqnarray}}

\begin{document}

\title{Protocol for reading out Majorana vortex qubit and testing non-Abelian statistics}

\author{Chun-Xiao Liu}
\affiliation{Kavli Institute for Theoretical Sciences, University of Chinese Academy of Sciences, Beijing 100190, China}
\affiliation{Condensed Matter Theory Center and Joint Quantum Institute,
Department of Physics, University of Maryland, College Park, MD 20742, USA}
\affiliation{QuTech and Kavli Institute of Nanoscience, Delft University of Technology, P.O. Box 4056, 2600 GA Delft, The Netherlands}

\author{Dong E. Liu}
\affiliation{State Key Laboratory of Low-Dimensional Quantum Physics and Department of Physics, Tsinghua University, Beijing 100084, China}
\affiliation{Frontier Science Center for Quantum Information, Beijing 100184, China}

\author{Fu-Chun Zhang}
\affiliation{Kavli Institute for Theoretical Sciences, University of Chinese Academy of Sciences, Beijing 100190, China}
\affiliation{CAS Center for Excellence in Topological Quantum Computation,
University of Chinese Academy of Science, Beijing, 100190, China}
\affiliation{Collaborative Innovation Center of Advanced Microstructures, Nanjing University, Nanjing 210093, China}

\author{Ching-Kai Chiu}\email{Corresponding: qiujingkai@ucas.edu.cn}
\affiliation{Kavli Institute for Theoretical Sciences, University of Chinese Academy of Sciences, Beijing 100190, China}
\affiliation{Condensed Matter Theory Center and Joint Quantum Institute,
Department of Physics, University of Maryland, College Park, MD 20742, USA}

\date{\today}

\begin{abstract}
The successful test of non-Abelian statistics not only serves as a milestone in fundamental physics but also provides a quantum gate operation in topological quantum computation. An accurate and efficient readout scheme of a topological qubit is an essential step toward the experimental confirmation of non-Abelian statistics. In the current work, we propose a protocol to read out the quantum state of a Majorana vortex qubit on a topological superconductor island. The protocol consists of four Majorana zero modes trapped in spatially well-separated vortex cores on  the two-dimensional surface of a Coulomb blockaded topological superconductor. Our proposed measurement is implemented by a pair of weakly coupled Majorana modes separately in touch with two normal metal leads, and the readout is realized by observing the conductance peak location in terms of gate voltage. Using this protocol, we can further test the non-Abelian statistics of Majorana zero modes in the two-dimensional platform. A successful readout of Majorana qubit is a crucial step towards the future application of topological quantum computation. In addition, this Coulomb blockaded setup can distinguish Majorana zero modes from Caroli-de Gennes-Matricon modes in vortex cores.
\end{abstract}

\maketitle

%

\section{introduction}

Majorana zero modes (MZM)~\cite{Read2000Paired, Kitaev2001Unpaired} are zero-energy quasiparticle excitations with neutral charge in the defects of the topological superconductors (TSC)~\cite{Nayak2008Non-Abelian, Alicea2012New, Leijnse2012Introduction, Beenakker2013Search, Stanescu2013Majorana, Jiang2013Non, DasSarma2015Majorana, Elliott2015Colloquium, Sato2016Majorana, Sato2017Topological, Aguado2017Majorana, Lutchyn2018Majorana}.  A pair of spatially separated MZMs forms a fermionic state whose number occupancy encodes the quantum information nonlocally. Such information is robust against most local perturbations and thereby is expected to possess much longer coherence time~\cite{KitaevTQC97}. Moreover, MZMs are non-Abelian anyons, i.e., exchanging two MZMs will rotate the quantum state in the degenerate subspace~\cite{KitaevTQC97,Ivanov2001NonAbelian}. Thus, quantum gates can be implemented by braid operations. Due to these two properties (nonlocal quantum information storage and non-Abelian statistics), MZMs are a promising candidate for realizing topological quantum computation.

Following multiple seminal theoretical proposals~\cite{Fu2008Superconducting, Sau2010Generic, Lutchyn2010Majorana, Oreg2010Helical, Zhang2018Helical}, much experimental progress has been made in the realization and detection of MZM within the context of both one-dimensional~\cite{Mourik2012Signatures,Rokhinson2012Fractional,Deng2012Anomalous,Das2012Zero,Churchill2013Superconductor,Finck2013Anomalous,Nadj-Perge14,Albrecht2016Exponential,Deng2016Majorana,Zhang2017Ballistic,ZhangNature2018} and  two-dimensional (2D) platforms~\cite{Xu2015Experimental, Sun2016Majorana,Wang2018Evidence,Zhang2018Observation,2018arXiv181208995M, Liu2018Robust,Gray2019Evidence, Wang2019Signature}. Regarding the 2D case, an $s$-wave superconducting surface with a single Dirac cone, forming an equivalent 2D $p \pm ip$ superconductor, is able to host an MZM emerging in the vortex core~\cite{Fu2008Superconducting}. The first candidate of this kind is the heterostructure of the topological insulator (${\rm Bi}_2{\rm Te}_3$) and the $s$-wave superconductor (${\rm NbSe}_2$). Superconducting proximity effect induces an $s$-wave SC pairing on the surface of the topological insulator. Recent scanning tunneling spectroscopy (STS) measurement on ${\rm Bi}_2 {\rm Se}_3$/${\rm NbSe}_2$ observed spin-polarization dependent zero-bias conductance peaks in the vortex cores of the heterostructure~\cite{Xu2015Experimental, Sun2016Majorana}, which are consistent with the MZM interpretation in tunnel spectroscopy~\cite{Law2009Majorana, Hu2016Theory}. However, a major practical problem is that other low-energy modes, e.g., Caroli-de Gennes-Matricon (CdGM) modes are very close to the zero mode so that they cannot be distinguished from the MZM~\cite{CAROLI1964307}. Recently, experiments on iron-based superconductors (${\rm FeTe}_{x}{\rm Se}_{1-x}$, ${\rm Li}_{1-x}{\rm Fe}_x {\rm OHFeSe}$) provide more evidence for the possible existence of MZMs in 2D systems, since the level spacing of those CdGM modes is larger than the STS resolution~\cite{Wang2018Evidence,Zhang2018Observation,2018arXiv181208995M, Liu2018Robust}. In addition, it has been experimentally observed\cite{conductance_plateau,Quantinzed_Chen_2019} that the tunneling conductance plateaux in the vortex cores are close to $2e^2/h$ at zero bias voltage. Given such encouraging experimental progresses made in the detection of the MZMs in vortex cores, naturally, the next milestone will be the readout of the quantum information encoded in MZMs. However, so far most of the experimental efforts are focusing on the Majorana resonance~\cite{Law2009Majorana}, which does not acquire quantum state information encoded. In addition, practical theoretical proposals for reading out MZMs in 2D TSC are still elusive, which limits the potential of demonstrating non-Abelian statistics stemming from MZMs in 2D systems.


In this work, we propose a theoretical scheme for reading out a Majorana vortex qubit (MVQ) and testing non-Abelian statistics. The MVQ consists of four MZMs trapped in spatially well-separated vortex cores on the 2D surface of a TSC island with finite charging energy, which could lead to Fu's Majorana teleportation~\cite{Fu2010Electron}. Our proposed setup, which is distinct from Fu's proposal, does not require a loop-geometry coherent channel and a threading flux to achieve electron interference, but includes an additional pair of Majorana modes. This island with four MZMs forms a minimum Majorana qubit, removing the extra structure complexity, and therefore more suitable and feasible for 2D superconducting island. Interestingly, the electron cotunneling process is identical to the Fu's teleportation case. Projective measurement can be implemented by a pair of weakly coupled Majorana modes in touch with separate normal metal leads in the Coulomb blockade valley, while high resolution readout of the measurement outcome can be realized by observing the conductance peak location at the charge degenerate points. This readout scheme can be used for braiding readout and further extended to Majorana fusion processes to demonstrate the non-Abelian statistics. The success of reading out Majorana qubit is a key step towards the future application of topological quantum computation. The rest of the paper is organized as follows. In Sec.~\ref{sec:readout}, which is the main part of the work, we introduce the theoretical model of Majorana vortex qubit and explain how to readout MVQ by a three-step scheme. Section~\ref{sec:test} shows the readout outcome for manifesting the non-Abelian nature of MZMs with the assumption that fusion and braiding operations can be successfully implemented in the TSC island. In Sec.~\ref{sec: moving}, possible experimental methods for moving vortices with MZMs are briefly discussed with an estimation of the time scale constraints.  In Sec.~\ref{sec:CdGM}, using the same experimental setup, we show how to distinguish between MZM and CdGM through tunnel conductance spectroscopy. Finally, Sec.~\ref{sec:summary} concludes the work with discussions.

\begin{figure}[t]
\begin{center}
\includegraphics[width=\linewidth]{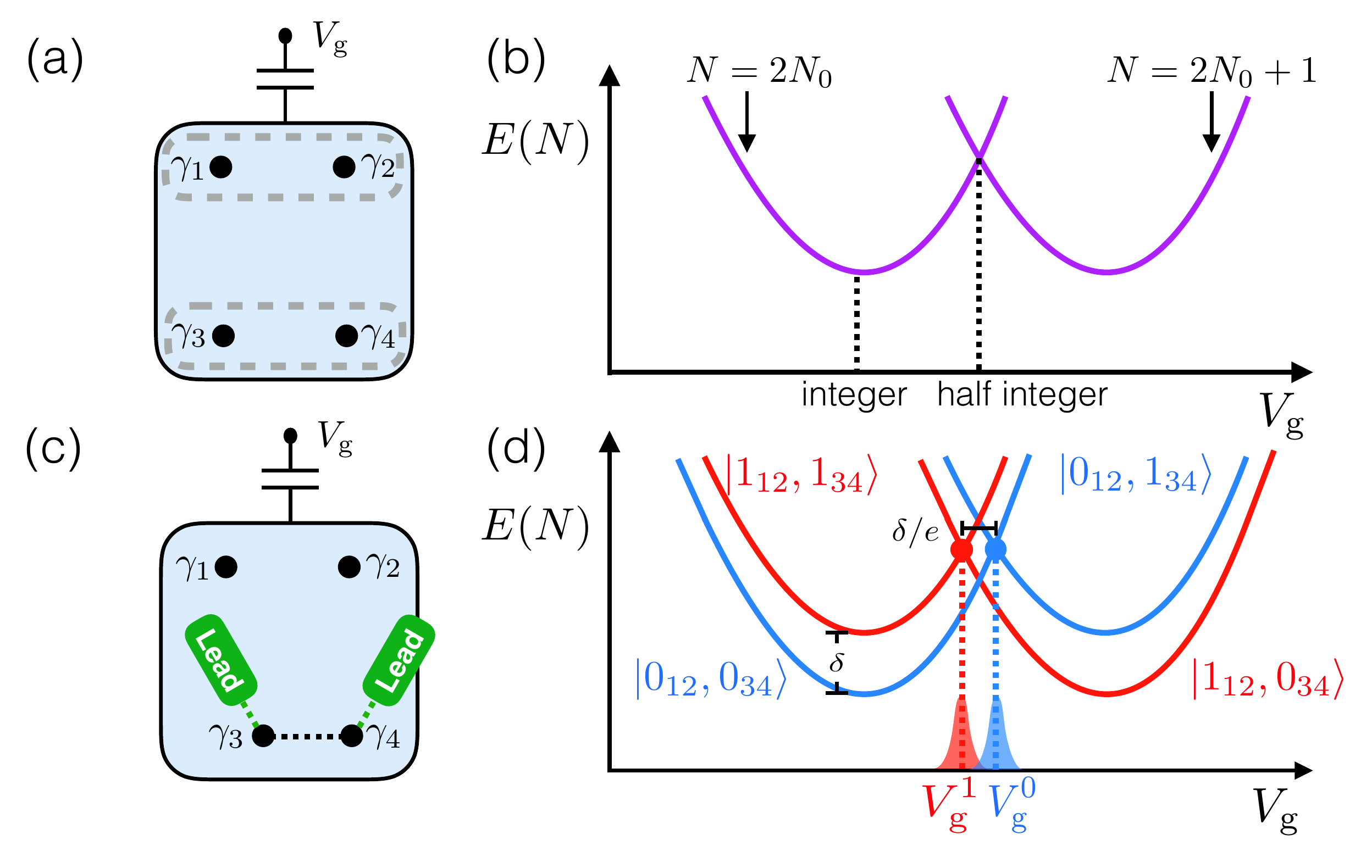}
\caption{(a) Schematic of a Majorana qubit. The Majorana qubit is a TSC island, which hosts four spatially separate MZMs and is capacitively connected to an adjustable external gate. (b) Energy spectrum of the TSC island as a function of the gate voltage $V_{\rm g}$. Each parabola corresponds to a fixed total electron number. (c) Schematic of the projective measurement. We first move $\gamma_3$ closer to $\gamma_4$, and then couple the two Majoranas to separate normal metal leads. (d) Readout process. Depending on the measurement outcome in (c), a conductance peak will show up either at $V^1_{\rm g}$(red peak) or at $V^0_{\rm g}$(blue peak). }
\label{fig:qubit} 
\end{center}
\end{figure}



\section{Readout scheme for Majorana vortex qubit}\label{sec:readout}

The MVQ is a TSC island hosting four MZMs trapped in spatially separate vortices, and the island is capacitively connected to an adjustable external gate, as shown in Fig.~\ref{fig:qubit}(a). The low-energy effective Hamiltonian of the TSC island is 
\begin{align}
H_C = E_C(N - N_{\rm g})^2 + n_{12}E_{12} + n_{34}E_{34},
\label{eq:H_C}
\end{align}
where $E_C = e^2/C$ is the charging energy, $N$ the total number of electrons in the TSC island, $N_{\rm g} = CV_{\rm g}/e$ the induced charge number tunable via the gate voltage ( in the following, we use $N_{\rm g}$ to represent the gate voltage). Among the four MZMs, as shown in Fig.~\ref{fig:qubit}(a), $\gamma_1$ and $\gamma_2$ ($\gamma_3$ and $\gamma_4$) pair up to form a normal fermionic state with energy $E_{12}$ ($E_{34}$) and occupancy $n_{12}$ ($n_{34}$). In the low-energy Hamiltonian of Eq.~\eqref{eq:H_C}, we assume MZMs to be the only relevant excitations, which is true in the low temperature limit $k_BT \ll \Delta$. The presence of charging energy constrains the Hilbert space: the parity of the total electron number should be equal to $ N~{\rm mod}~2 = (n_{12} + n_{34} )~{\rm mod}~2  $. If the four MZMs are spatially well-separated ($E_{12}, E_{34} \to 0$) and the gate voltage is tuned near an integer point $N_{\rm g} \approx 2N_0$, as shown in Fig.~\ref{fig:qubit}(b), there will be two degenerate ground states in the TSC island: $\ket{2N_0; 0_{12}, 0_{34}}$ and $\ket{2N_0; 1_{12}, 1_{34}}$. These two states span the degenerate subspace of a MVQ. In the Coulomb blockaded valley, the MVQ has the protection from the spatially separate MZMs, and in addition, the charging energy can significantly reduce the notorious ``quasi-particle poisoning'' from outside environment~\cite{Plugge2017Majorana, Karzig2017Scalable, Pikulin2019Coulomb}.

 The readout of an arbitrary MVQ state $\ket{\psi} = a \ket{0_{12}, 0_{34}} + b \ket{1_{12}, 1_{34}} $ requires a sequence of efficient projective measurements to obtain the probabilities $|a|^2$ and $|b|^2$. In our scheme, the projective measurement is implemented by measuring the two-terminal conductance via a pair of weakly coupled Majorana modes on the TSC island. Here we outline the three-step protocol for a single readout process:

(i) For a TSC island in the Coulomb blockade valley $N_{\rm g} \approx 2N_0$, we adiabatically lift the double degeneracy between $\ket{2N_0; 0_{12}, 0_{34}} $ and $\ket{2N_0;  1_{12}, 1_{34}} $ by moving $\gamma_3$ closer to $\gamma_4$ ($E_{34} >0$). 

(ii) We weakly couple the two Majorana modes $\gamma_3$ and $\gamma_4$ to separate normal metal leads near the vortex cores, as shown in Fig.~\ref{fig:qubit}(c). The two-terminal conductance measurement would project the MVQ in one of the basis states after a measurement time $t\sim(\Gamma_L \Gamma_R V/E_C^2)^{-1}$. The tunnel broadening of the MZM is defined as $\Gamma_{L(R)} = 2 \pi |t_{L(R)}|^2 \rho$, with $\rho$ being the lead density of states and $t_{L(R)}$ being the coupling strength between the MZM and left (right) lead. $V$ is the voltage drop between the source and drain leads.

(iii) We tune the gate voltage toward the half integer point $N_{\rm g} \approx 2N_0+1/2$. Depending on the measurement outcome in step ii, a conductance peak will show up at the charge degenerate point either to the left or to the right of the half integer point [Fig.~\ref{fig:qubit}(d)]. This peak location tells us in which basis state the MVQ is projected.

The three steps mentioned above lead to a single readout of a MVQ state. To have a statistical estimate of $|a|^2$ and $|b|^2$, we need to repeat the operation for many times: prepare the qubit into the same initial state and then perform the readout procedure. Now, we try to flesh out the three-step protocol:

The goal of \emph{step i} is to lift the double degeneracy between the two ground states $\ket{2N_0;  0_{12}, 0_{34}} $ and $\ket{2N_0;  1_{12}, 1_{34}} $. Initially, the four MZMs are spatially well-separated ($E_{12}, E_{34} \to 0$) and the SC island is in the Coulomb blockade valley $N_{\rm g} \approx 2N_0$. We then adiabatically move $\gamma_3$ closer to $\gamma_4$, while still keeping $\gamma_{1,2}$ faraway from each other and from $\gamma_{3,4}$. More details of the vortex movement will be discussed in Sec.~\ref{sec: moving}.
Due to the wavefunction overlap between $\gamma_3$ and $\gamma_4$, the coupling between the two modes becomes finite~\cite{Cheng2009Splitting,PhysRevB.82.094504}:
\begin{align}
E_{34} \approx \frac{\cos \Big( p_F R_{34} + {\pi}/{4} \Big)}{\sqrt{p_F R_{34}}} \exp \Big( -R_{34}/\xi \Big),
\end{align}
in the large distance limit $R_{34} \gg \xi$. Here, $R_{34}$ is the intervortex distance between $\gamma_3$ and $\gamma_4$, $\xi$ the SC coherence length, and  $p_F$ the Fermi momentum of the surface Dirac cone. In the limit of $p_F \rightarrow 0$, the energy does not oscillate, and we can safely assume $E_{34} = \delta >0$ to hold when we adjust $R_{34}$. Now the energies of the two lowest eigenstates are split: $E(2N_0; 1_{12}, 1_{34}) - E(2N_0; 0_{12}, 0_{34}) = \delta$, and this positive energy splitting is crucial to the readout measurement in step iii. We emphasize that the energy splitting cannot lead to any parity change of the Fermion state of $\gamma_{3,4}$ since the Fermion parity in the entire system is protected by the charge energy and $\gamma_{3,4}$ do not couple well-separated $\gamma_1$ or $\gamma_2$. 

In \emph{step ii}, we perform a projective measurement on the MVQ state by a two-terminal conductance measurement in the Coulomb blockade valley. We weakly couple the Majorana modes $\gamma_{3,4}$ to separate normal metal leads. This can be realized either by two normal electrodes, two STM tips, or one electrode and one tip~\cite{Dvir2018Tunneling, Dvir2018Spectroscopy, Pang2015Majorana, Xi2015Ising, Ge2015Development, Kubo2006Epitaxially}. For the STM tip, the tunneling strength between the tip and the Majorana modes inside the vortex cores is determined by the spatial separation between the tip and the vortex core. For the normal electrode case, a dielectric layer needs to be placed between the normal electrode and the vortex core. The corresponding tunneling Hamiltonian is~\cite{Fu2010Electron}
\begin{align}
H_T &\approx -t_L d_L \gamma_3 e^{i\phi/2} -t_R d_R \gamma_4 e^{i\phi/2} + {\rm H.c.},
\end{align}
where $t_{L(R)}$ is the effective coupling between the left (right) lead and Majorana $\gamma_{3(4)}$, $d_{L(R)}$ annihilates one electron in the left (right) lead, $e^{i\phi/2}$ increases the number of electrons in the SC island by one, and. For the SC island in the Coulomb blockade valley ($N_{\rm g} \approx 2N_0$), the dominant process for the electron tunneling between the source-drain leads is the single electron cotunneling process, which is a second-order process in terms of the tunneling Hamiltonian~\cite{supp}:
\begin{align}
&H_{\rm co} = \sum_m \frac{ \langle 2N_0; n_{34} | H_T \ket{ m}  \langle m | H_T \ket{2N_0; n_{34}} }{E(m) - E(2N_0; n_{34})} \nn
&\approx \frac{-2i t_L t^*_R }{E_C} \Bigg[ \frac{2n_{34}-1}{1 - 4(\Delta N)^2} + \frac{\delta}{E_C} \frac{1 + 4(\Delta N)^2}{\left( 1 - 4(\Delta N)^2 \right)^2} \Bigg] d^{\dagger}_R d_L + {\rm H.c.},
\label{eq:H_co}
\end{align}
where $\Delta N = N_{\rm g} - 2N_0$, and $\ket{ m}$ is a virtual state with total number of electrons $m=2N_0 \pm 1$. The leading-order term in the cotunneling transmission amplitude has a $\pi$-phase shift depending on the occupancy $n_{34}$, while the next-order correction is a constant of order $O(\delta/E_C)$. During the whole process, $n_{12}$ is safely protected by the large Majorana separation. Note that Eq.~\eqref{eq:H_co} is equivalent to the transmission Hamiltonian derived in Refs.~\cite{Fu2010Electron, Plugge2017Majorana}, but instead of a loop-geometry coherent channel and a threading flux used for nanowire systems with two Majorana modes, here we use four Majorana modes to achieve the same coherent electron transport. The elimination of a loop geometry makes our proposal more feasible to 2D superconducting island.
Now we turn on a small bias voltage between the two leads. After a sufficient long time $t > \big( \Gamma_L \Gamma_R V / E^2_C \big)^{-1}$, the electric current flows through $\gamma_1$ and $\gamma_2$ so that the conductance measurement becomes projective, i.e., the MVQ state $\ket{\psi} = a\ket{ 0_{12}, 0_{34}} + b\ket{ 1_{12}, 1_{34}}$ will collapse into either $\ket{ 0_{12}, 0_{34}}$ with probability $|a|^2$, or $\ket{1_{12}, 1_{34}}$ with probability $|b|^2 = 1 - |a|^2$~\cite{Plugge2017Majorana}. However, the difference between the conductance magnitude for the two basis states with $n_{34}=0$ or 1 is tiny, which is of order $O(\delta/E_C)$ as shown in Eq.~\eqref{eq:H_co}. Thereby even though the MVQ is projected into one of the basis states, it is hard to tell which one by merely observing the cotunneling conductance magnitude.

Naturally, the goal of \emph{step iii} is to read out the outcome of the projective measurement in a more transparent way. The key is to tune the gate voltage close to the half integer point $N_{\rm g} \approx 2N_0 + 1/2$, which is near the charge degenerate point. At the charge degenerate point of the two states with the same $n_{12}$ occupancy, a conductance peak arises because an electron can freely tunnel in and out of the Majorana modes $\gamma_3$ and $\gamma_4$  without costing additional energy. The reason for fixed $n_{12}$ in the projected state is that isolated $\gamma_1$ and $\gamma_2$ are always topologically protected. Due to the energy splitting introduced in \emph{step i}, as shown in Fig.~\ref{fig:qubit}(d), the charge degenerate point for $\ket{2N_0; 0_{12}, 0_{34}}$ and $\ket{2N_0+1; 0_{12}, 1_{34}}$ is shifted to the right of the half integer point (blue dot), located at $N^{0}_{\rm g} = 2N_0+ 1/2 + \delta/2E_C$, while that for $\ket{2N_0; 1_{12}, 1_{34}}$ and $\ket{2N_0+1; 1_{12}, 0_{34}}$ is shifted to the left of the half integer point (red dot), located at $N^{1}_{\rm g} = 2N_0+ 1/2 - \delta/2E_C$. Therefore, after the MVQ is projected in $\ket{0_{12},0_{34}}$ ($\ket{1_{12},1_{34}}$)  in the Coulomb blockade valley, as we tune the gate voltage toward half integer point, a blue (red) peak will show up at $V^0_{\rm g}$ ($V^1_{\rm g}$).  In other words, observing the conductance peak location in gate voltage completes the readout of a MVQ. Furthermore, in order to have high resolution in the readout measurement, the tunneling and thermal broadening need to be less than the conductance peak separation ($\Gamma, k_BT < \delta$). 

On the other hand, the three-step protocol could serve as the initialization of a qubit. Once the target initialized state is confirmed by the conductance peak location, we tune the gate voltage back to the Coulomb blockade valley, remove the normal metal leads, and separate the two Majorana vortices. Consequently, this qubit state is initialized and ready for quantum information processing.


\begin{figure}[t]
\begin{center}
\includegraphics[width=\linewidth]{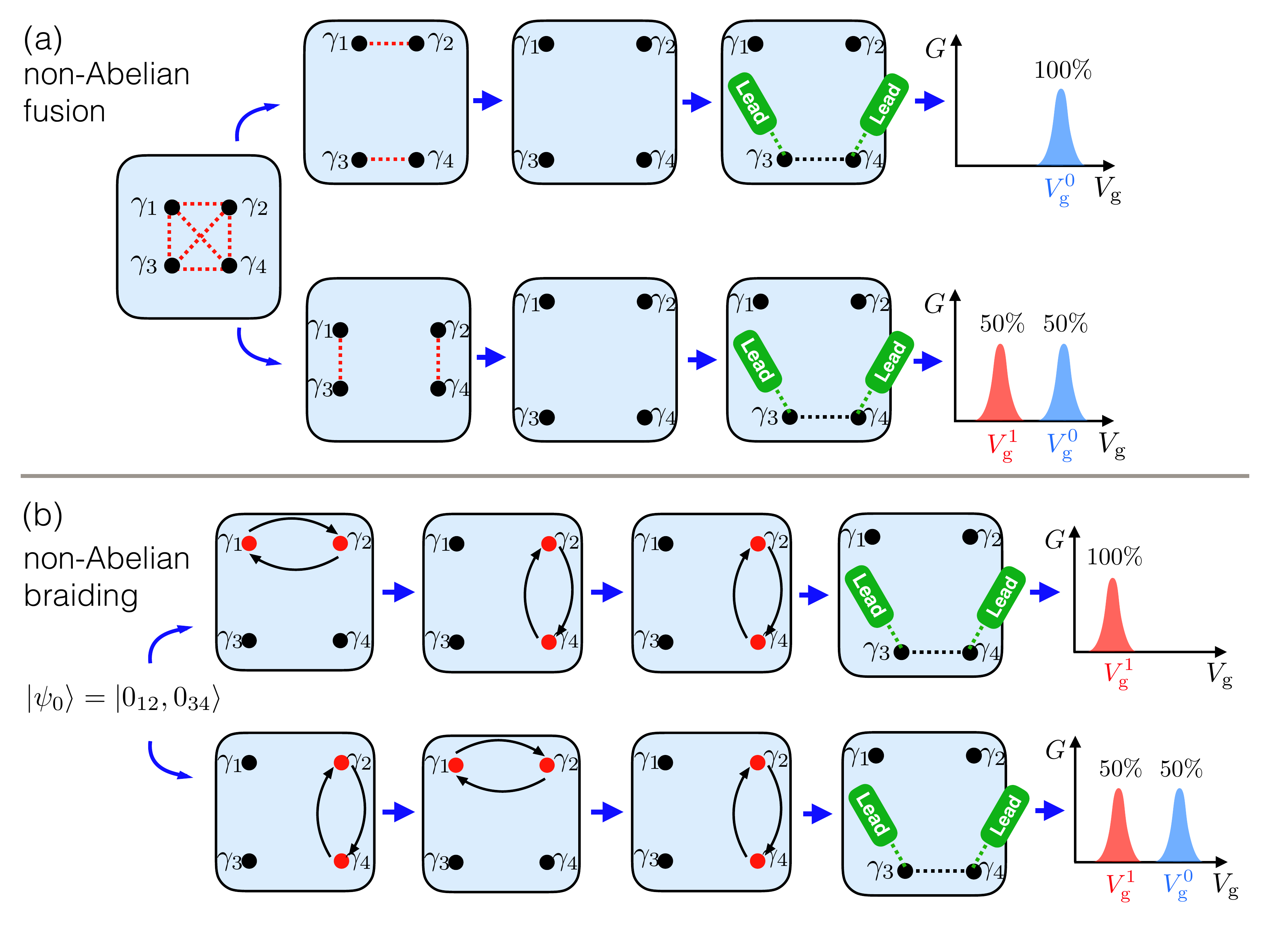}
\caption{(a) Proposal for testing the non-Abelian fusion of MZMs. Initially, all the four Majoranas are coupled with each other. Upper and lower paths correspond to different ways of adiabatically decoupling the Majoranas, leading to a different final state. (b)Proposal for testing the non-Abelian braiding of MZMs. Starting from the same initial state $\ket{0_{12}, 0_{34}}$, the qubit system is then applied by the same set of braid operations but in a different order. The final states corresponding to the two orders are distinct. }
\label{fig:nonAbelian} 
\end{center}
\end{figure}

\section{testing the non-Abelian nature of Majorana zero modes} \label{sec:test}

We now briefly discuss two proposals for testing the non-Abelian nature of MZMs -- one through the fusion process and the other through the braiding process. We emphasize that our focus is still on the readout of the final quantum states, assuming that the fusion and braiding processes can be successfully implemented. A detailed discussion of the experimental realization of fusing and braiding MZMs is beyond the scope of the current work. The schematic of the fusion proposal is shown in Fig.~\ref{fig:nonAbelian}(a). Initially four MZMs are spatially nearby and couple to each other, and the system equilibrates to the unique ground state. In the upper path, we adiabatically separate $\gamma_{1,2}$ from $\gamma_{3,4}$, and then move each MZM away from each other to suppress any MZM hybridization. This gives the final state $\ket{0_{12}, 0_{34}}$, since in the intermediate step, $n_{12}$ and $n_{34}$ are good quantum numbers~\cite{supp}. In the lower path, however, starting from the same initial state, we first separate $\gamma_{1,3}$ from $\gamma_{2,4}$ and then move each MZM away from each other. For the same reason, the corresponding final state would be $\ket{0_{13}, 0_{24}}= \frac{1}{\sqrt{2}} \big( |0_{12},0_{34} \rangle - i|1_{12},1_{34} \rangle \big)$. We thus see that two procedures to create MZMs in the vacuum lead to two different final states, which provides a direct demonstration of the non-Abelian fusion of MZMs~\cite{Nayak2008Non-Abelian}. The distinction between the two final states can be revealed by the conductance peak locations in our readout protocol -- the upper path results in a peak only located at $V^0_{\rm g}$, while the lower one leads to a peak located at $V^0_{\rm g}$ or $V^1_{\rm g}$ with equal probability.  On the other hand, the schematic for the non-Abelian braiding process is shown in Fig.~\ref{fig:nonAbelian}(b). Suppose the MVQ is initialized in state $\ket{0_{12},0_{34}}$ and  braid operation has been experimentally achieved. In the upper path, after we adiabatically exchange $\gamma_{1,2}$, and then exchange $\gamma_{2,4}$ for twice, the final state is given by $-e^{-i \pi/4}\ket{1_{12}, 1_{34}}$. In the lower path, however, starting from the same initial state, we first exchange $\gamma_{2,4}$, then $\gamma_{1,2}$, and then $\gamma_{2,4}$ again. The resulting final state would be $\frac{-i}{\sqrt{2}} \big( \ket{ 0_{12}, 0_{34}} -i\ket{ 1_{12}, 1_{34}}\big)$. The same set of braid operations, which is applied on the same initial state in a different order, leads to two distinct final states. This thereby demonstrates the non-Abelian braiding of MZMs. Similar to the fusion process, this distinction between two final states can be revealed by the location of the conductance peaks in our readout protocol.

\section{Moving vortices with MZMs}\label{sec: moving}

Eventually, the readout of the Majorana vortex modes requires tuning the mutual Majorana coupling strengths. The straightforward to achieve this coupling adjustment is to meticulously move vortices with MZMs. Although precisely controlling the locations of the vortices has been a difficult task, it is encouraging that 
experimentalists have been able to control the locations of vortices by using magnetic force microscopy in a thin film of superconducting Niobium\cite{doi:10.1063/1.3000963}. The recent proposal extends this technique to manipulate vortices in ${\rm FeTe}_{x}{\rm Se}_{1-x}$~\cite{November2019Scheme}. This is promising that controlling vortices with MZMs can be achieved in the near future. 



	${\rm FeTe}_{x}{\rm Se}_{1-x}$ is one of the ideal platforms hosting MZMs in vortex cores. We use ${\rm FeTe}_{x}{\rm Se}_{1-x}$ with the real physical parameters as an example to provide a practical recipe for vortex movement without poisoning the MVQ before the MVQ readout. We adapt the simulation program in [\onlinecite{2019arXiv190413374C}] faithfully describing Majorana physics on the surface of ${\rm FeTe}_{x}{\rm Se}_{1-x}$. At the beginning of the MVQ readout, we keep four MZMs residing in four vortices separately far apart with roughly $100$nm intervortex distance so that their suppressed hybridization protects the MVQ. To achieve the readout, we bring one of the vortices ($\gamma_3$) close to another ($\gamma_4$) as illustrated in Fig.~\ref{fig:qubit}(c). Since the hybridization energy $\Delta E$ is in order of $0.1\mu$eV in the simulation, we expect the time scale to move the vortex should be short enough ($\tau\ll \hbar/\Delta E=7$ns) so that the Majorana hybridization does not poison the MVQ. On the other hand, the lowest energy ($\Delta^2/E_F$) of the CdGM modes is close to the Majorana splitting $\delta=0.2$meV\cite{Chen:2018aa,2019arXiv190413374C,2018arXiv181208995M,2019arXiv190102293K}. The time scale of the Majorana movement should be long enough ($\tau \gg \frac{\hbar}{\Delta^2/E_F}=3$ps) so that the MZMs cannot be excited to the CdGM states.  On the other hand, since currently the best energy resolution of STM is around $20\mu$eV\cite{doi:10.1063/1.5049619}, to have detectable energy splitting we suggest the distance of the two hybridized MZMs be $40$nm, based on the Majorana physics simulation in ${\rm FeTe}_{x}{\rm Se}_{1-x}$\cite{2019arXiv190413374C}. Fortunately, $40$nm is experimentally reachable distance between the STM tips, since the two tips $72$nm apart were successfully made in the lab already\cite{Kubo2006Epitaxially}. For the readout, adjusting the Majorana distance from $100$nm to $40$nm is much more accessible than physically braiding two Majorana modes, since the braiding leads to twisted vortex lines in the bulk. We emphasize that this energy splitting does not affect the time scale of the vortex movement since the charge energy can protect the fermion parity of the two isolated hybridized MZMs ($\gamma_1$ and $\gamma_2$). This is the main idea of our proposal by using the protection of the charge energy.

	The non-Abelian fusion we propose can be achieved by using the similar length and time scales. For the non-Abelian braiding, similarly the time scale of MZM exchanging is in the time region of the readout. Alternatively, without exchanging the positions of vortex cores, exchanging Majorana modes can also be implemented through three successive projective measurements~\cite{Bonderson2008Measurement,Vijay2016Teleportation}.





\begin{figure}[tbp]
\begin{center}
\includegraphics[width=\linewidth]{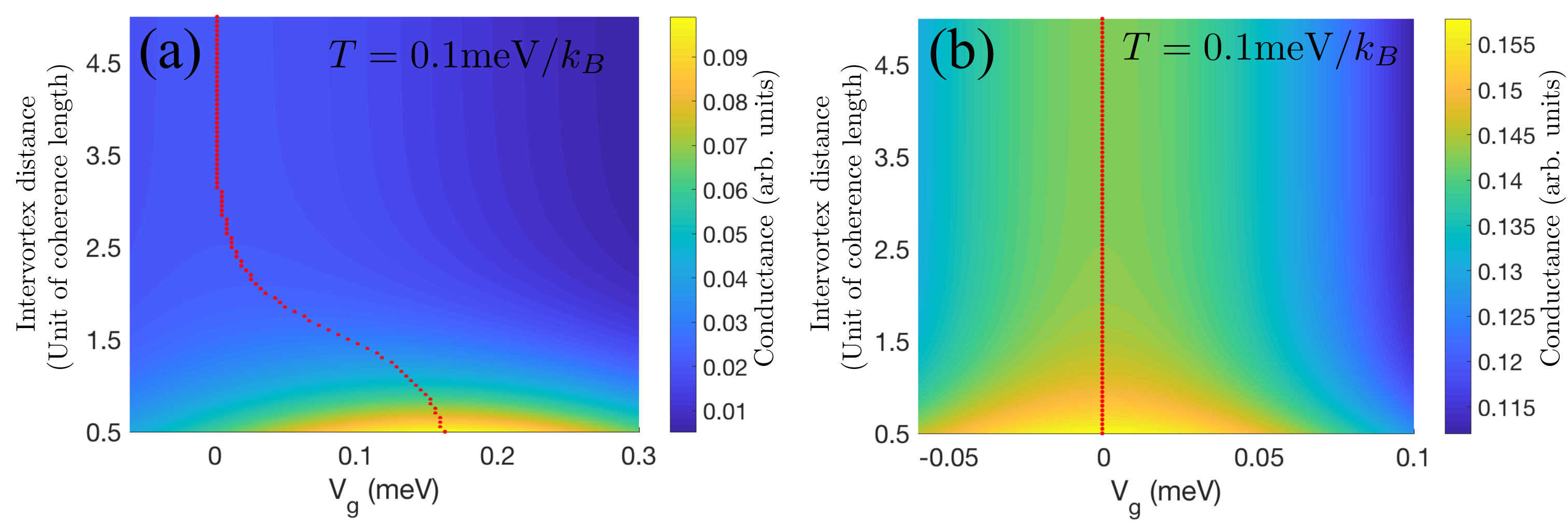}
\caption{Conductance for MZMs and CdGM modes as a function of gate voltage and intervortex distance. Here, $V_{\rm g}=0$ corresponds to the half integer point $N_{\rm g}=2N_0+1/2$ and the coherence length of the MZMs and the CdGM modes is chosen to be the unity. (a) Conductance for two separated CdGM modes of energies $0.2$ and $0.21$ meV in the absence of MZMs. At $k_BT \approx 0.2$ meV, the conductance peak location (red dashed line) shifts from $V_{\rm g} \approx 0.16$ mV to 0 mV as the intervortex distance increases. (b) Conductance for two MZMs in two separate vortex cores,  with a single CdGM mode of energy $0.2$ meV inside one of the vortices. The conductance peak always appears at $V_{\rm g}=0$ mV. 
}
\label{Conductance_CdGM_MZM} 
\end{center}
\end{figure}



\section{Caroli-de Gennes-Matricon modes}\label{sec:CdGM}

CdGM modes with low energy always coexist with a MZM in a vortex core~\cite{CAROLI1964307}. As we will show, near the charge energy degenerate point, the readout measurement through resonant tunneling not only accurately reads out the MVQ but also clearly distinguishes MZMs from CdGM modes when the temperature is very close to the energy of the CdGM modes.    


	We first consider the two vortex cores in contact with the leads possess only two CdGM modes separately with energies $E_1$ and $E_2$, where $E_1 \lesssim E_2$~\cite{supp}. When these two vortices are spatially close, the overlapping of the two CdGM modes leads to the gate voltage ($V_{\rm g}$) location of the conductance peak near $E_1$ and $E_2$ referenced to the half integer point, since the CdGM modes become effective extended states~\cite{conductance_coulomb_blockade_roman}. At $k_BT\sim E_1,\ E_2$, as the intervortex distance increases, the conductance monotonically decreases and the $V_{\rm g}$ location of the conductance peak moves to the half integer point~\cite{chiu_blockade}, which is identical to the case in the presence of the MZMs as shown in Fig.~\ref{Conductance_CdGM_MZM}(b). 
	
	On the other hand, consider each vortex possesses a MZM and one of the two vortices has a CdGM mode with energy $E_1>0$. At any temperature, the gate voltage of the conductance peak is located at zero due to the strong electron cotunneling assisted by the MZMs~\cite{Fu2010Electron}. For $k_BT\sim E_1$, the CdGM mode contributes a small portion of the conductance peak in the short intervortex distance. As the intervortex distance increases, the conductance from the CdGM contribution is suppressed. 
	

Thus, the key to distinguish MZM from CdGM mode is that at the temperature comparable to the energy of the CdGM mode, the conductance peak location stays (varies) in the presence (absence) of MZMs as the intervortex distance increases.  

\section{Summary and discussions}\label{sec:summary}

Strong clues of a single MZM bounded in a vortex core of the TSC has been revealed by the recent experiment~\cite{Wang2018Evidence,Zhang2018Observation,2018arXiv181208995M, Liu2018Robust}. Taking advantage of this important progress, we propose a feasible protocol for reading out the quantum information encoded in the vortex MZMs by using the well-developed Coulomb blockade transport measurement~\cite{Albrecht2016Exponential}. We expect the size of the ${\rm FeTe}_{x}{\rm Se}_{1-x}$ island to be close to the London penetration depth ($\sim 500$nm)~\cite{PhysRevB.81.180503,PhysRevB.82.184506} to host few MZMs; hence, this small island leads to large charging energy ($<2.9$meV)~\cite{supp}, which is greater than experimental temperature, protecting the MVQ. This is encouraging that similar superconducting islands hosting few vortices were made in the lab\cite{SERRIERGARCIA2017109}.
Since the readout of the MVQ always requires the hybridization of the two MZMs, the quantum information of the original qubit might be lost due to the quasiparticle poisoning and the relaxation to the ground state.  The two isolated MZMs $\gamma_{1,2}$ and the Coulomb blockaded superconductor with large charging energy protects the MVQ during the readout progress. 
The true value of the tunneling conductance, which is commonly affected by the unknown realistic details, is difficult to be a suitable observable to determine the quantum state. To circumvent this problem, we show that the $V_{\rm g}$ location of the conductance peak can determine the eigenstate of MVQ. This is the key idea of our readout protocol -- regardless of the precise conductance value, measuring the $V_{\rm g}$ location of the conductance peak is more experimentally feasible in principle; this is compared to original Aharonov-Bohm interferometer~\cite{Fu2010Electron} which requires strong coherence and an extra metallic arm in the device. Furthermore, the extension of this readout protocol can be applied to Majorana qubits in nanowire setups~\cite{Karzig2017Scalable, Plugge2017Majorana}.  

The tunneling of the Coulomb blockaded island can further confirm the presence of the MZMs with the coexistence of the CdGM modes. 
At temperature comparable to  the energy of the CdGM modes, the gate voltage location of the peak conductance is an important observable again to distinguish the MZMs from the CdGM modes.
Confirming the existence of the MZMs in the vortices is the primary step to experimentally achieve our readout protocol.  

Another important implementation of the Coulomb blockade design is the initialization of the quantum state protected by charging energy. Using this initialization, we create multiple identical quantum states (disregarding phase difference) to statistically test non-Abelian statistics. Experimental demonstration of Majorana non-Abelian statistics is a milestone toward topological quantum computation. Our proposal paves a path to reach this goal.




We thank H. Ding, M. Franz, D.L. Feng, T. Hanaguri, T.Y. Liu, J. Jia, L.Y. Kong, D.F. Wang, T. Machida, S. Das Sarma, R. Wiesendanger, H. Zhang, and H. Zheng for fruitful discussions. C.-X.L. and C.-K.C. acknowledge the support by Laboratory for Physical Sciences and Microsoft. C.-X. L. is supported by a subsidy for top consortia for knowledge and innovation (TKl toeslag) by the Dutch ministry of economic affairs. C.-K.C. and F.-C.Z. are supported by the Strategic Priority Research Program of the Chinese Academy of Sciences (Grant XDB28000000). D.E.L is supported by the State Key Laboratory of Low-Dimensional Quantum Physics at Tsinghua University. F.-C.Z. is also supported by the National Science Foundation of China (Grant NSFC-11674278).

\appendix

\section{Derivation of cotunneling Hamiltonian}
Here we are going to derive the low-energy Hamiltonian for the single-electron cotunneling process between two normal metal leads. Majorana modes $\gamma_{3,4}$ inside the vortex cores of SC island are weakly coupled to each other possibly due to wavefunction overlap, and in addition we weakly couple $\gamma_{3,4}$ with separate normal metal leads. The corresponding Hamiltonian is
\begin{align}
&H_{\rm SC} = E_C(N - N_{\rm g})^2 +  n_{34} \delta, \nonumber \\
&H_{\rm leads} = \sum_{\alpha = L/R,k} \xi_{\alpha,k} c^{\dagger}_{\alpha,k}c_{\alpha,k} \nonumber \\
&H_T = -t_L d_L \gamma_3 e^{i\phi/2} -t_R d_R \gamma_4 e^{i\phi/2} + {\rm H.c.}. 
\end{align}
$H_{\rm SC}$ is the Hamiltonian of the TSC island with total electron number $N$ and charging energy $E_C$, $N_{\rm g}$ is the gate electron number, $\delta$ the strength of coupling between $\gamma_{3}$ and $\gamma_4$, and $n_{34}$ the occupation number of the normal fermion state composed by $\gamma_3$ and $\gamma_4$. $H_{\rm leads}$ is the Hamiltonian of the two separate normal metal leads, $c_{\alpha,k}$ is the electron in lead-$\alpha$, and $\xi_{\alpha,k}$ is the occupation energy. $d_{\alpha} = \sum_k c_{\alpha,k}$ is the electron at the contact point to the vortex in the TSC island. $H_T$ is the coupling Hamiltonian between the lead electrons and the Majorana modes inside the vortex core. $t_{\alpha}$ is the effective coupling strength between the lead electron and the Majorana mode, and $e^{i\phi/2}$ increases the total number of electrons inside the SC island by one. We assume that the gate voltage is tuned in the Coulomb blockade valley ($N_{\rm g} \approx 2N_0$) such that the transmission of electrons through the island is dominated by a second-order process. Here we only consider the transmission process of an electron tunneling from the left lead to the right lead (the opposite process from right to the left lead is simply the Hermitian conjugate term). For state $\ket{2N_0;0_{12},0_{34}}$, the effective Hamiltonian is
\begin{widetext}	
\begin{align}
H_{00} = & \frac{ \bra{2N_0;00} -t^*_R d^{\dagger}_R \gamma_4 e^{-i\phi/2}  \ket{2N_0+1;01} \bra{2N_0+1;01} -t_L d_L \gamma_3 e^{i\phi/2} \ket{2N_0;00} }{E(2N_0+1;01) - E(2N_0;00)} \nonumber \\
&+ \frac{ \bra{2N_0;00}  -t_L d_L \gamma_3 e^{i\phi/2}  \ket{2N_0-1;01} \bra{2N_0-1;01} -t^*_R d^{\dagger}_R \gamma_4 e^{-i\phi/2} \ket{2N_0;00} }{E(2N_0-1;01) - E(2N_0;00)} \nonumber \\
= & \frac{-2i t_L t^*_R d^{\dagger}_R d_L }{E_C} \frac{1 + \delta/E_C}{ (1+\delta/E_C)^2 - 4 \Delta N^2 },
\end{align}	
where $ \Delta N = N_{\rm g} - 2N_0$. Note that the virtual state in the second-order process $\ket{2N_0 \pm 1;01}$ can hold one more or one less electron with the fermion state by $\gamma_{3,4}$ being occupied. Similarly for state $\ket{2N_0;1_{12},1_{34}}$, we will get	
\begin{align}
H_{11} = & \frac{ \bra{2N_0;11} -t^*_R d^{\dagger}_R \gamma_4 e^{-i\phi/2}  \ket{2N_0+1;10} \bra{2N_0+1;10} -t_L d_L \gamma_3 e^{i\phi/2} \ket{2N_0;11} }{E(2N_0+1;10) - E(2N_0;11)} \nonumber \\
&+ \frac{ \bra{2N_0;11}  -t_L d_L \gamma_3 e^{i\phi/2}  \ket{2N_0-1;10} \bra{2N_0-1;10} -t^*_R d^{\dagger}_R \gamma_4 e^{-i\phi/2} \ket{2N_0;11} }{E(2N_0-1;10) - E(2N_0;11)} \nonumber \\
= & \frac{2i t_L t^*_R d^{\dagger}_R d_L }{E_C} \frac{1 - \delta/E_C}{ (1-\delta/E_C)^2 - 4 \Delta N^2 }.
\end{align}
\end{widetext}	
Compared to $H_{00}$, we notice that the virtual state in the second-order process $\ket{2N_0 \pm 1;10}$ has the fermion state by $\gamma_{3,4}$ be unoccupied, which explains why $\delta \to -\delta$. What's more, there is an additional overall factor of $-1$ relative to $H_{00}$ due to the opposite parity of the ground state. Note that there is no off-diagonal terms like $H_{01}$ or $H_{10}$ because the ground state fermi parity does not change after a single electron tunneling for twice. Thus the cotunneling Hamiltonian including $H_{00}$ and $H_{11}$ is
\begin{align}
H_{co} &= \frac{-2i t_L t^*_R d^{\dagger}_R d_L }{E_C} \frac{ 1 + \delta/E_C S}{(1 + \delta/E_C S)^2 - 4 \Delta N^2 }S \nonumber \\
& \approx \frac{-2i t_L t^*_R d^{\dagger}_R d_L }{E_C} \Bigg( \frac{1}{1-4 \Delta N^2}S - \frac{\delta}{E_C} \frac{1+4\Delta N^2}{( 1 - 4\Delta N^2 )^2} \Bigg),
\end{align} 
where we only keep terms up to $O(\delta/E_C)$, and $S$ is a prefactor with $S=1$ for $\ket{2N_0;00}$ and $S=-1$ for $\ket{2N_0;11}$.

\begin{figure}[tbp]
\begin{center}
\includegraphics[width=0.4\linewidth]{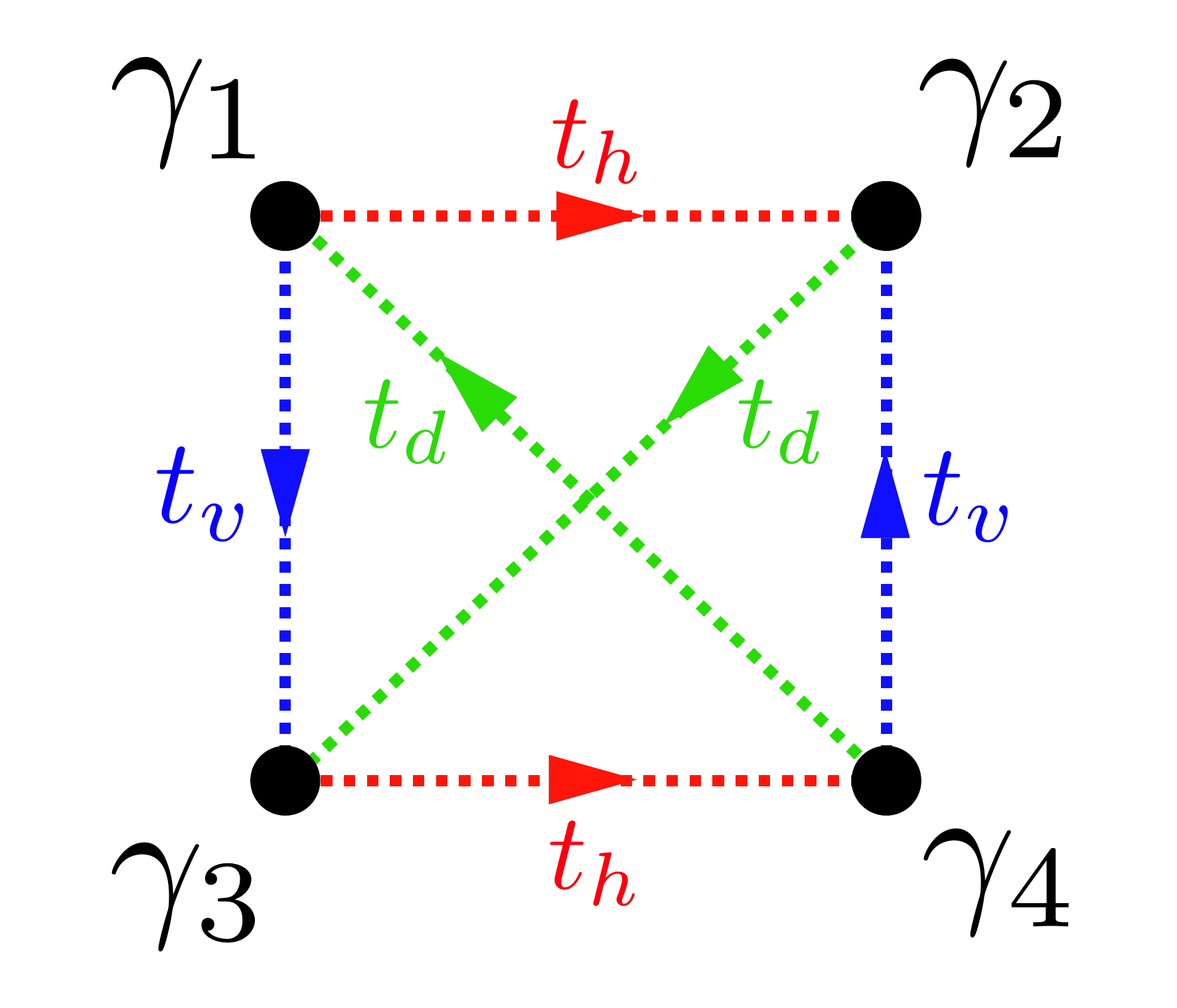}
\caption{Four weakly coupled Majorana modes. The coupling amplitude between each pair of Majorana modes is proportional to the strength of mutual coupling $t_{\alpha}$ and a phase factor $s_{mn} = \pm i$. The sign of the phase factor is constrained by the $Z_2$ gauge symmetry of the magnetic field, and is indicated by the arrow here: coupling along the arrow direction gives $s=i$, while coupling opposite to the arrow direction gives $s=-i$.}
\label{fig:4MZM} 
\end{center}
\end{figure}

\section{fusion process for four Majorana modes on a SC island}
For the fusion process, our goal is to show that for four initially mutually coupled Majorana modes, different ways/paths of separating all them apart lead to different ground states. Such path dependence of the final ground state is a direct demonstration of the non-Abelian nature of MZMs (i.e., Ising anyons). The Hamiltonian for the mutually coupled four Majorana modes on a SC island is
\begin{align}
&H = H_{h} + H_{v} + H_{d} + H_C \nonumber \\
&H_h = it_h(\gamma_1 \gamma_2 + \gamma_3 \gamma_4  ) = -2t_h ( \tau_z \oplus 0) \nonumber \\
&H_v = it_v(\gamma_1 \gamma_3 - \gamma_2 \gamma_4  ) = 2t_v ( \tau_y \oplus 0) \nonumber \\
&H_d = it_d( -\gamma_1 \gamma_4 + \gamma_2 \gamma_3  ) = -2t_d (0 \oplus \sigma_x) \nonumber \\
&H_C = E_C (0 \oplus \sigma_0).
\end{align}
For each pair of coupled Majorana modes in the form of $s t_{mn}\gamma_m \gamma_n$, $t_{mn}$ denotes the strength of coupling, which we set as a positive value in the low chemical potential limit, and $s=\pm i$ is a phase factor constrained by the $Z_2$ gauge symmetry of the magnetic field~\cite{PhysRevB.92.134519}. Here we define $f_1 = ( \gamma_1 + i \gamma_2 )/2$ and $f_2 = ( \gamma_3 + i \gamma_4 )/2$, and the Hilbert space is spanned by $( \ket{0}, f^{\dagger}_1 f^{\dagger}_2 \ket{0}, f^{\dagger}_1 \ket{0},  f^{\dagger}_2 \ket{0} )$. Due to the parity symmetry in the coupling Hamiltonian, each term of the coupling Hamiltonian can be diagonalized into two blocks with even and odd parity respectively. $\tau_{\alpha}$ ($\alpha=0,x,y,z$) are Pauli matrices acting on the even parity subspace spanned by $( \ket{0}, f^{\dagger}_1 f^{\dagger}_2 \ket{0} )$, while $\sigma_{\alpha}$ are Pauli matrices acting on the odd parity subspace spanned by $( f^{\dagger}_1 \ket{0},  f^{\dagger}_2 \ket{0} )$, and $0$ is a null matrix. In the strong Coulomb blockade limit $E_C \gg t_{mn}$, the subspace with odd parity is pushed upward in energy, and we can simply project the total Hamiltonian onto the even parity subspace in the calculation of the ground state. The projected Hamiltonian and its eigenvalues are
\begin{align}
&H' = -2t_h \tau_z + 2t_v \tau_y \nonumber \\
&E = \pm 2 \sqrt{ t^2_h + t^2_v }.
\end{align}
Initially when the four Majorana modes are coupled with $t_h = t_v$, the ground state is $\ket{\Psi} \propto ( 1 + \sqrt{2}, i )$. If we adiabatically turn off the coupling $t_v$ while keep $t_h$ unchanged as shown in the upper path of fusion panel in Fig.~(2), the ground state will adiabatically evolve into $\ket{\Psi} = ( 1 , 0 ) = \ket{0}$. On the other hand, if we adiabatically turn off the coupling $t_h$ while keep $t_v$ unchanged as shown in the lower path of fusion panel in Fig.~(2), the final ground state will be $\ket{\Psi} = ( 1 , -i ) \propto  \ket{0} - i f^{\dagger}_1 f^{\dagger}_2 \ket{0}$. Note that there is no level crossing between the ground state and the first excited state in either of the adiabatic process, as the level crossing requires all the coefficients of Pauli matrices to vanish, which does not happen in the middle of the adiabatic process we consider here. Therefore the ground state is uniquely determined. We thus have shown that different ways of separating weakly coupled Majorana modes lead to different final ground states. Although the non-Abelian ``separation'' process of MZMs is opposite to the ``fusion'' process, it can equally demonstrate the non-Abelian statistics of MZMs.

\section{Distinguishing MZM from CdGM modes}

\begin{figure}[t]
\begin{center}
\includegraphics[clip,width=0.20\columnwidth]{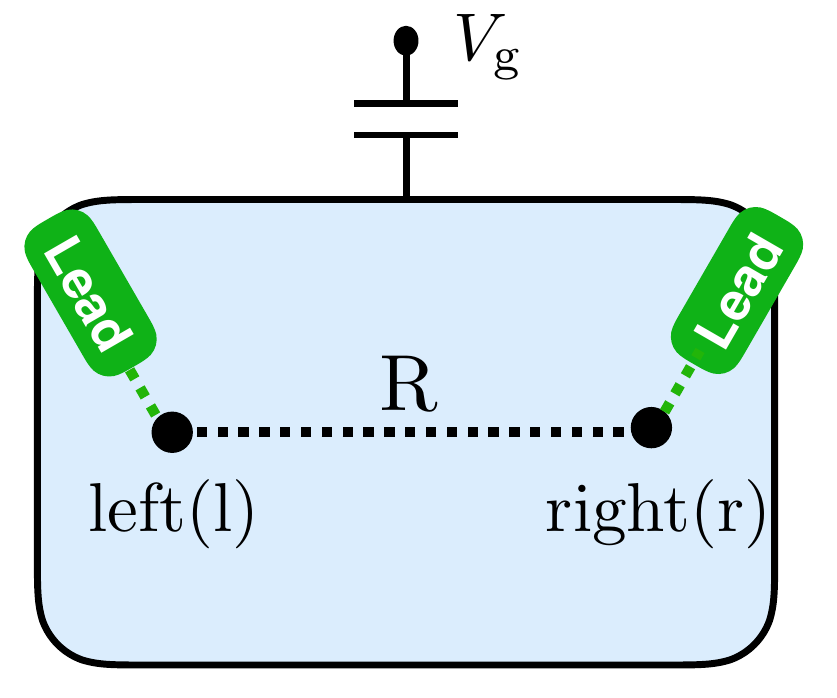}
\caption{The two lead tips weakly couple two vortices separately on the surface of the superconductor. 
 }
\label{MZM_CdGM_Schematic} 
\end{center}
\end{figure}

	Measuring the two-terminal conductance of the Coulomb-blockaded superconducting island can provide the readout of Majorana qubits. Our readout proposal motivates us to check if the location of the conductance peak can differentiate the MZMs and CdGM modes. Hence, we focus on the gate voltage ($V_g$) near resonant tunneling and use the master equation of the superconducting Coulomb-blockade~\cite{chiu_blockade} to capture the physics of the weak tunneling. Consider two vortices trapping MZMs or CdGM modes and two lead tips are moved to weakly couple the two vortex cores separately  as the two terminals for the tunneling as shown in Fig.~\ref{MZM_CdGM_Schematic}. We are interested in two cases --- (a) each of the vortices possesses one CdGM mode only (b) one vortex has one MZM only and the other has one MZM and one CdGM mode. Therefore, the tunneling physics can be effectively described by two fermions with energy $E_i$ ($i=1, 2$) and the tunneling rates for these two fermions are labelled by $\Gamma_i^{l,r}$ for particle tunneling in the left/right vortex and  $\Lambda_i^{l,r}$ for hole tunneling in the left/right vortex. To simplify the problem, we consider the resonant tunneling between the electron numbers $N$ and $N-1$, where $N$ is even. Since at temperature $T$ in equilibrium the general form of the conductance for the superconducting island via two low-energy Fermions was derived in~\cite{chiu_blockade} (refer to equations 2.26 $-$ 2.30), we simply use the formula 2.30 in \cite{chiu_blockade}
\begin{widetext}	
\begin{small}
\begin{align}
\frac{dI}{dV}=&\beta e^2 \bigg \{ \peq(0,1) f(\epsilon_1)\frac{\Gamma^l_1\Gamma_1^r}{\Gamma_1} +\peq(1,0) f(\epsilon_2)\frac{\Gamma^l_2\Gamma_2^r}{\Gamma_2}+\peq (1,0) f(
\tilde{\epsilon}_1)\frac{\Lambda^l_1\Lambda_1^r}{\Lambda_1}+\peq(0,1) f(\tilde{\epsilon}_2)\frac{\Lambda^l_2\Lambda_2^r}{\Lambda_2} \nonumber \\
&- (\gamma^l_1-\gamma_2^l+\lambda^l_1-\lambda_2^l)(\gamma_1^r-\gamma_2^r+\lambda_1^r-\lambda_2^r) \nonumber \\
&\times \Big( \frac{1}{\peq (0,1) f(\epsilon_1)\Gamma_1}+\frac{1}{\peq (1,0) f(\epsilon_2)\Gamma_2}+\frac{1}{\peq (1,0) f(\tilde{\epsilon}_1)\Lambda_1}+\frac{1}{\peq (0,1) f(\tilde{\epsilon}_2)\Lambda_2} \Big )^{-1} \bigg \}, \label{odd_even}
\end{align}
\end{small}
\noindent where $\Gamma_i=\Gamma_i^l+\Gamma_i^r$, $\Lambda_i=\Lambda_i^l+\Lambda_i^r$, $\gamma_i^\alpha=\Gamma_i^\alpha/\Gamma$, $\lambda_i^\alpha=\Lambda_i^\alpha/\Lambda$ ($\alpha=l,r$), $\beta=1/k_BT$. The complicated tunneling formula is determined by various functions, which are the Gibbs distributions $\peq(1,0)=e^{-\beta E_1}/Z$, $\peq(0,1)=e^{-\beta E_2}/Z$, and the Fermi-Dirac distribution $f(\epsilon)=1/(1+e^{\beta \epsilon})$, where $Z=e^{-\beta E_1}+e^{-\beta E_2}+E^{-\beta(E_1+E_2+\Delta U)}+e^{-\beta\Delta U}$, $\Delta U=E_c(2N-2N_g-1)$, $\epsilon_i=E_i+\Delta U$, $\tilde{\epsilon}_i=-E_i+\Delta U$. Once the physical values of $E_1, E_2, T$, and the tunneling rates are given, the conductance formula is a function of the gate voltage $V_g$, where $V_g=eN_g/C$. We change the reference of the gate voltage $V_g\rightarrow V_g - \frac{C(2N-1)}{2e}$ so that the charging energies with the electron number $N-1$ and $N$ are identical ($\Delta U=0$) at $V_g=0$, which is the conductance peak of the resonant tunneling via MZMs~\cite{Fu_teleportation}. 
\end{widetext}

(a) While each vortex possesses one CdGM mode, the spatial distribution of the CdGM wavefunction $\Psi(r)$ can be captured by the Bessel functions~\cite{PhysRevLett.80.2921,PhysRevLett.115.177001}. With the coherence length $\xi$ of the superconductor, the wavefunction is approximately proportional to $\Psi(r) \sim e^{-r/\xi}$ with Friedel-like oscillation, of which the period is given by Fermi wavelength $1/k_F$ from bulk or surface band at the Fermi level. Here we consider $k_F=0$ so that the wavefunction exhibits only the spatial exponential decay. For the CdGM mode with energy $E_1$ located at the left vortex, we can assume the tunneling rates $\Gamma_1^l=\Lambda_1^l=1$ and $\Gamma_1^r=\Lambda_1^r=e^{-2R/\xi}$, where $R$ is the intervortex distance; the exponential decay of the tunneling rates in the right vortex core stems from the spatial distribution of the CdGM mode trapped in the left vortex. Similarly, for the CdGM mode with $E_2$ located at the right vortex, the tunneling rates are given by $\Gamma_2^r=\Lambda_2^r=1$ and $\Gamma_2^l=\Lambda_2^l=e^{-2R/\xi}$. We note that knowing the exact value of the conductance is not required in this case since it will be shown later that the location of the conductance peak is the main observable to distinguish MZMs and CdGM modes. We choose $E_1=0.2$meV and $E_2=0.21$meV and use eq.~\ref{odd_even} to compute the conductance at $T=0.1$meV$\sim E_1,E_2$ and $T=0.01$meV$\ll E_1, E_2$. 

	When these two vortices are spatially close, the overlapping of the two CdGM modes leads to the strong tunneling conductance and the gate voltage ($V_g$) of the conductance peak is near $E_1$ and $E_2$, since the CdGM modes become effective extended states~\cite{conductance_coulomb_blockade_roman}. 
	For $k_BT\sim E_1,\ E_2$, when the distance ($R$) of the two vortices increases, the conductance monotonically decreases and is close to a non-zero constant, and the gate voltage of the conductance peak moves to zero, which is identical to the resonant point in the presence of the MZMs as shown in Fig.~3 (a) and Fig.~\ref{Conductance_CdGM_MZM} (c). It has been studied~\cite{chiu_blockade} that two separate localized fermion vortex modes can have the non-zero conductance and the conductance peak located at $V_g=0$. This tunneling stems from the thermal fluctuation of the two fermions. In the low-temperature limit ($k_BT\ll E_1,\ E_2$), for any intervortex distance, the gate voltage of the conductance peak is always located at $E_1$ unless $R\gg 1$. However, the conductance peak exponentially decreases to zero when the intervortex distance increases as shown in Fig.~\ref{Conductance_CdGM_MZM}(b). The reason is that when the tunneling of the thermal fluctuation for the localized fermions is suppressed, the conductance is approximately proportional to $e^{-2R/\xi}$ from the exponential wavefunction decay of the CdGM mode.  

	(b) while each vortex possesses an MZM, the left vortex traps an additional CdGM mode with energy $E_1$. On the one hand, the tunneling rates of the CdGM mode are identical to case (a) ($\Gamma_1^l=\Lambda_1^l=1$ and $\Gamma_1^r=\Lambda_1^r=e^{-2R/\xi}$). On the other hand, we consider MZM wavefunctions $\Phi_l(r)$ and $\Phi_r(r)$ are located on the left and right vortices respectively. With the spatial exponential decay, $|\Phi_l(r_l)|\gg|\Phi_l(r_r)|$ and $|\Phi_r(r_r)|\gg|\Phi_r(r_l)|$, where $r_l$ and $r_r$ are the locations of the left and right vortices. Since the two MZMs form a fermion with $E_2=0$ and the tunneling rates of this fermion state at the two different vortex cores are proportional to $|\Phi_l(r_l)|^2$ and $|\Phi_r(r_r)|^2$ separately, the tunneling rates are independent of the intervortex distance $R$. Therefore, without loss of generality we can assume the tunneling rates of the MZMs $\Gamma_2^r=\Lambda_2^r=\Gamma_2^l=\Lambda_2^l=1$. Similarly, the conductance as a function of $V_g$ is computed at $T=0.1$meV$\sim E_1,\ E_2$ and $T=0.01$meV$\ll E_1, E_2$.

In the presence of the MZMs trapped by the two vortices, the conductance peak behaves differently from the vortices without MZMs. At any temperature, the gate voltage of the conductance peak is located at zero and the conductance is never suppressed in any circumstances due to the resonant tunneling via the MZMs~\cite{Fu_teleportation}. In other words, the tunneling rates at the two terminals are non-zero. For $k_BT\sim E_1$, the CdGM mode contributes a small portion of the conductance peak in the short intervortex distance as shown in Fig.~3(b) and Fig.~\ref{Conductance_CdGM_MZM}(d). As the intervortex distance increases, the CdGM contribution is suppressed and the conductance peak is reduced and stays at a non-zero constant. On the other hand, for $k_BT\ll E_1$, the conductance peak stemming from only the resonant tunneling via MZMs is always a non-zero constant at any intervortex distance as shown in Fig.~\ref{Conductance_CdGM_MZM}(b,d).
%
%
%
%
%
%

\begin{figure}[t!]
\begin{center}
\includegraphics[clip,width=0.95\columnwidth]{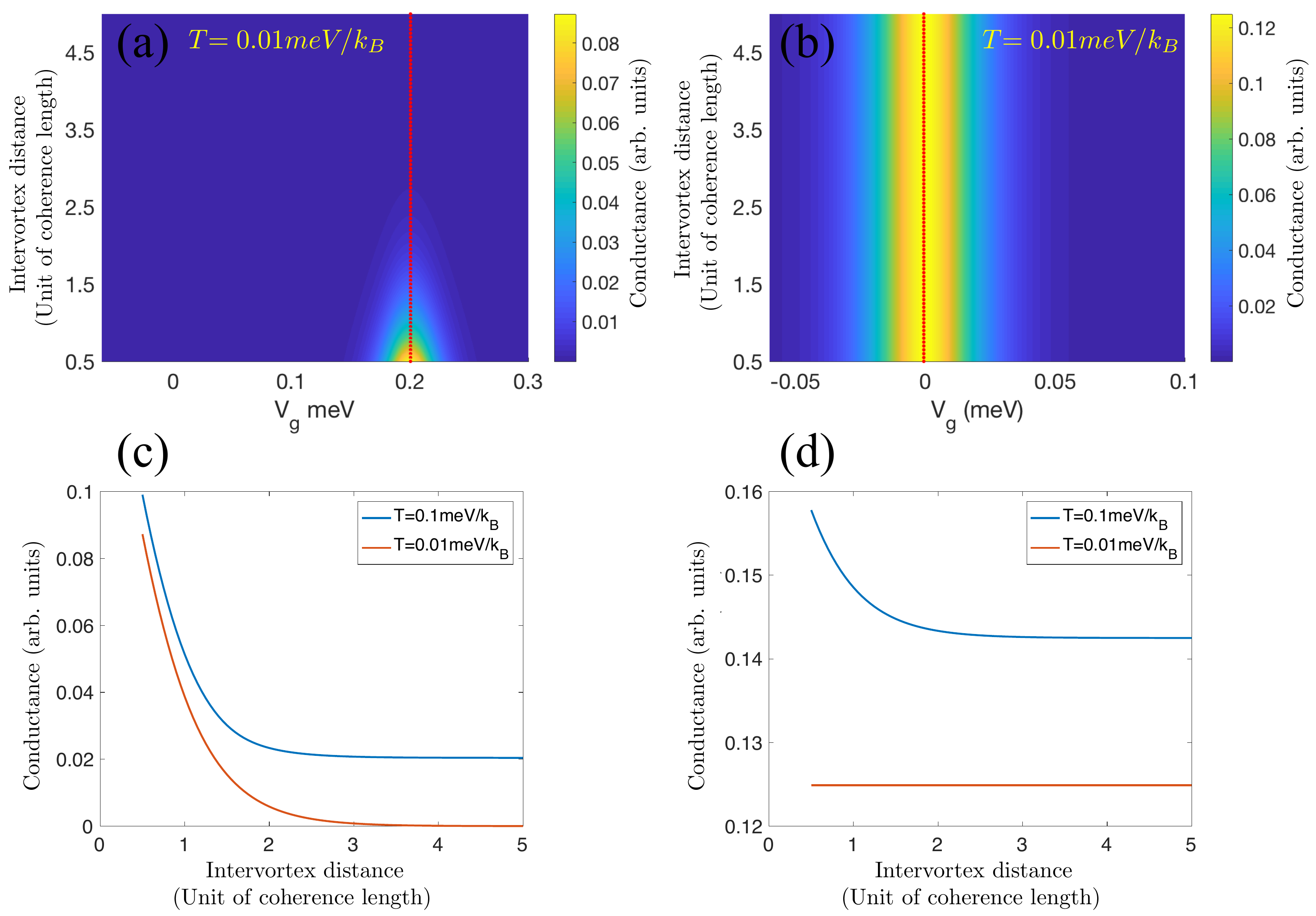}
\caption{The tunneling conductance (a,b) and the conductance peak (c,d) of the Coulomb blockaded superconducting island with the two leads in contact with the two vortices respectively, where $V_g=0$ corresponds to the resonant point $N_g=N+1/2$. (a,c) two CdGM modes located at two vortex cores spatially separated have $0.2$ and $0.21$ meV energies respectively without MZMs. Since the red dashed lines in (a,b) represent the conductance peaks, for $k_BT\ll E_1$, the $V_g$ location of the conductance peak is independent of the intervortex distance. Panel (c) shows the conductance exponentially decays to zero as a function of the intervortex distance at $k_BT\ll E_1$. (b,d) two MZMs are located at two vortices separately and one of the two vortices possesses a CdGM mode with $0.2$meV energy. The conductance mainly stemming from the resonant tunneling via the MZMs is always a non-zero constant at any intervortex distance. For $k_BT\sim E_1$ and the short intervortex distance, the non-zero conductance stems from the CdGM mode and the MZMs. For the long intervortex distance, due to the suppression of the CdGM conductance the MZMs lead to the non-zero constant conductance. 
 }
\label{Conductance_CdGM_MZM} 
\end{center}
\end{figure}

\section{charging energy estimation}

The charging energy estimation of the superconducting island is an important requirement to determine the feasibility of our readout proposal. First, the surface diameter of the island must be greater than London penetration depth to collect enough magnetic flux for the formation of Abrikosov vortices. Second, the island has to be thick enough to avoid the coupling of the two Dirac cones on the top and bottom surfaces, since the MZMs vanish in the vortices once the Dirac cones disappear. The decay length of this cone coupling is approximately given by $v_F/\delta=2nm$, where the Fermi velocity of the Dirac cone $v_F\approx 20 $nm$\cdot$meV and the bulk gap connecting the surface Dirac cone $\delta\sim10$meV~\cite{Wang2018Evidence}. However, $2$nm is not thick enough. When the thickness ($d$) of the superconductor is smaller than the London penetration depth of the bulk material ($\lambda$), the effective London penetration depth for the thin film depends on the thickness ($\lambda_{\rm{eff}}=\lambda^2/d$)~\cite{deGennesbook}. That is, a thinner SC film leads to longer London penetration depth. Therefore, to have the localized Dirac surface and the minimum of the London penetration depth, the thickness of the SC island should be greater than the original London penetration depth ($\sim$500nm~\cite{PhysRevB.82.184506,PhysRevB.81.180503}). 

	We can roughly estimate the charging energy of the island by considering a sphere with radius $r=500$nm
\bee
E_c=\frac{e^2}{C}=\frac{e^2}{4\pi\epsilon_{r}\epsilon_0 r}=2.9\rm{meV},
\ee
where we use $\epsilon_r=1$ in the estimation. However, the substrate always possesses the relative permittivity $\epsilon_r$ greater than one, the charging energy of the island is commonly less than $2.9$meV. 
%


\bibliography{BibMajorana,TOPO}


\end{document}